\documentclass[preprint,review,12pt]{elsarticle}
\usepackage{sectsty}

\usepackage[margin=2.5cm]{geometry}
\usepackage{setspace}

\usepackage{relsize}
\usepackage{titlecaps}

\usepackage{titlesec}

\titleformat*{\section}{\MakeUppercase}
\titleformat*{\subsection}{\bfseries}
\titleformat*{\subsubsection}{\itshape}
\usepackage{float}
\usepackage{multirow}
\usepackage{textcomp}
\usepackage{graphicx}
\usepackage{amsmath}
\usepackage{longtable}
\usepackage{textcomp}
\usepackage{bm}
\usepackage{paralist}
\usepackage{natbib}
\journal{The Canadian Journal of Chemical Engineering}

\biboptions{numbers,sort&compress}
\biboptions{super}
\biboptions{square}

\begin{document}

\begin{frontmatter}
\title{Understanding the influence of rheological properties of shear-thinning liquids on segmented flow in microchannel using CLSVOF based CFD model}
\author{Somasekhara Goud Sontti}
\author{Pankaj G. Pallewar}
\author{Amritendu Bhuson Ghosh}
\author{Arnab Atta\corref{cor1}}
\cortext[cor1]{Corresponding author. Tel.: +91 3222 283910}
\ead{arnab@che.iitkgp.ac.in}
\address{Multiscale Computational Fluid Dynamics (mCFD) Laboratory, Department of Chemical Engineering, Indian Institute of Technology Kharagpur, West Bengal 721302, India}

\begin{abstract}
\begin{spacing}{1}	
\noindent In this study, two phase gas\textendash shear\textendash thinning liquid flow in square microchannel is numerically investigated using coupled level set and volume of fluid (CLSVOF) method. Systematic investigation is carried out to explore the influence of Polyacrylamide (PAM) concentration, surface tension, velocity ratios, and contact angle on the gas slug length, volume, unit cell length, and pressure drop. Three different concentrations of PAM solutions, which exhibit shear-thinning behaviour are considered as the continuous phase. Gas slug length, volume, and unit cell length decreased with increasing the PAM concentration. Velocity and non-homogeneous viscosity distributions in the liquid slug for three different PAM concentration solutions are reported. Gas slug length decreases with an increase of the contact angle and the bubble shape change from convex to concave. This numerical work provides the fundamental insights in segmented flow formation and two\textendash phase flow characteristics comprising shear\textendash thinning liquids.
\end{spacing}
\end{abstract}


\end{frontmatter}

\section*{Introduction}

Over the last decade, microfluidics has gained considerable attention in a wide range of academic investigations and industrial applications.\cite{anna-2003, whitesides2006origins} Microfluidic devices can be used in different areas like mass and heat transfer processes, as well as in reactions e.g., in a reactor \cite{li-2014}, absorber \cite{ganapathy-2014,shao2010mass}, desorber \cite{liu-2016}, distillation unit\cite{liu-2016}, and also synthesis of micro\textendash and nanoparticles.\citep{li-2010,li-2011} Depending on the fluid properties and microchannel geometric configurations, different flow patterns such as bubble, slug, churn, annular, or stratified flows are observed. Numerous researchers have reported different flow regime maps for two\textendash phase systems.\citep{akbar-2002,rebrov2010two,triplett-1999} Different geometric configurations such as co\textendash flow\citep{gupta2010,wang-2015,haase-2016,haase2017}, T\textendash junction\citep{garstecki-2006,kawahara-2011,santos-2010}, and flow\textendash focusing devises\cite{weber2007computational,dietrich2008bubble,fu-2010} have been proposed over the years for segmented flow formation and analysis. Taylor bubble flow is observed to be typically surrounded by a thin liquid film adjacent to the wall, and two consecutive gas bubbles are separated by a liquid slug. In case of hydrophobic channel walls, the gas phase occupies the entire cross\textendash section of channels without any surrounding film thickness, which can be termed as segmented flow to differentiate from Taylor bubble flow.\cite{Fletcher2017} 

\noindent For gold and silver nanoparticles synthesis, Baber et al., \cite{baber2017engineering} demonstrated the potential of using microfluidic reactor with a coaxial flow in controlling nanoparticle size and dispersity. Experimental studies on segmented flow have shown that it was highly influential in the synthesis of nanoparticles due to its flexibility in manipulation of mass transfer and hydrodynamics characteristics.\cite{kohler2013micro} Furthermore, with the incorporation of segmented flow for the gold nanoparticle synthesis, it has been observed that axial dispersion of the slugs is enhanced, which in turn improved the interaction of the gold nuclei. \cite{sebastian2012size} Abolhasani et al., \cite{abolhasani2015oscillatory} studied the bi\textendash phasic droplet or segmented flow dynamics of the Palladium catalysed C\textendash C and C\textendash N coupling reactions based on sequential input of the aqueous and organic phases. They pointed out that due to differences in the surface energies of the phases involved, mass transfer rate and mixing characteristics were improved in T\textendash junction microchannel. Kinetic study of palladium nano\textendash rod synthesis in segmented flow regime suggested that residence time distributions were narrower, which reduced the raw material costs, and mass transfer rates were found to be enhanced, thus promoting anisotropic growth of the nanocrystals. \cite{sebastian2016continuous} Wang et al., \cite{wang2017liquid} studied segmented flow for a range of reaction schemes such as $H_2$$O_2$ oxidation, etherification of benzyl bromide and phenoxide, and C\textendash C and C\textendash N coupling reactions. They reported higher conversion and better yields resulting from improved reaction control, faster mass transfer and mixing, which provide a platform for integrating reaction intensification. Pan et al., \cite{pan2017controllable} showed that gas\textendash liquid segmented flow could be used for commercial scale-up of multi-step reactions. Some researchers have numerically reported Taylor flow behaviour without any liquid film\cite{kumar-2007}, while few acknowledged the limitation of not capturing film thickness due to poor grid resolutions.\citep{ganapathy-2013,qian-2006,yu-2007} 

\noindent Interestingly, most of the previous research were concerned with bubble formation in two-phase gas\textendash Newtonian liquid systems, while several fluids involved in numerous application are likely to exhibit non\textendash Newtonian behaviour \citep{arratia-2008,groisman-2003,li-1999,nghe-2011}. Yang et al., \cite{yang-2010} reported experimental observations of flow regimes using a gas\textendash shear\textendash thinning liquid system, which demonstrated significant influences of rheological properties of the fluid on flow regime transition. Mansour et al., \cite{manso-2015} compared gas slug length, velocity, and pressure drop for both Newtonian and non\textendash Newtonian liquids. Considerable effect on flow parameters were experienced due to rheological properties. Fu et al., \cite{fu-2011} investigated bubble formation mechanism and effect of flow rates on hydrodynamic parameters using shear\textendash thinning liquids in a T\textendash junction. Chen et al., \cite{chen-2013} developed a three-dimensional numerical model for bubble formation in a T\textendash junction microchannel in Newtonian and non\textendash Newtonian liquids using volume-of-fluid (VOF) method. Initially the developed model is verified for Newtonian fluids with in\textendash house experimental visualization, and subsequently the study was extended for power\textendash law and Bingham fluids.

\noindent From the aforementioned discussion, it can be recognized that most of the previous studies have utilized the standard geometry for segmented flow formation. However, lower pressure drop at the junction is anticipated in converging (or Y-type) configuration as compared to cross pattern\citep{dang-2013}. 

\noindent Recently our group reported the viscous effects on Taylor bubble formation by considering power-law viscosity model in a co-flow microchannel \citep{sontti2017cfd,sontti2018formation}. We also presented flow pattern maps in a co-flow microchannel based on gas and liquid phase inlet velocities for power-law liquids \citep{sontti2017numerical}. Moreover, it can also be realized that as compared to Newtonian liquids, understanding of segmented flow in microfluidic devices for non\textendash Newtonian liquids is still at an elementary stage. Therefore, in this work, we computationally examine the influence of rheological properties of shear\textendash thinning liquids on the formation of segmented flow, slug length and volume, and pressure drop. Systematic studies are carried out to realize the effect of inlet velocity ratios, surface tension, and wall adhesion properties on segmented flow behaviour in converging shape microchannel.


 
 
\section*{Modeling approach}

To capture the fluid-fluid interface, VOF method\citep{hirt1981} is widely used owing to its low computational cost and ease of implementation.\citep{guo-2014,gupta-2010} However, in low Capillary number flows standard VOF methods often suffers from a drawback related to spurious currents\cite{harvie2006} arising from surface tension modelling. Few researchers \cite{popinet2003gerris, guo2015, francois2006balanced, denner2014fully} have improved the balanced VOF model with more accurate surface tension modelling. Level set (LS) method\citep{suss-1994} is capable of tracking the interface accurately however, it induces loss of mass conservation.\citep{suss-1994,sussman2000} To circumvent the inadequacies of both approaches, a coupled LS and VOF (CLSVOF) method is implemented in this work, where the interface normal and curvature are calculated using the LS function. Exact position of the interface is adjusted by balancing the volume in each cell so that the volume fraction value calculated from VOF is satisfied.

\subsection*{Volume of Fluid (VOF) Method}
In VOF approach, for solving the governing equations for immiscible fluids, following set of assumptions is considered:

\begin{compactitem}
	\item The fluids in both phases are in incompressible
	\item Flow regime is laminar  
	\item	The two-phase flow is isothermal
	\item	The microchannel is completely wet
	\item	No mixing happens between the two phases
	\item	No velocity-slip at the walls
	\item	No thermophysical property variation with temperature	
\end{compactitem}


\noindent \textbf{Equation of continuity:} 

\begin{equation}
\label{eq:mass_eqn}
\frac{\partial  \rho }{\partial t}  +  \nabla .  ( \rho  \vec{ U } ) =0
\end{equation}

\noindent \textbf{Equation of motion:}

\begin{equation}
\label{eq:mom_eqn}
\frac{ \partial (\rho \vec{ U })}{ \partial t} + \nabla.( \rho \vec{ U } \vec{ U }) = - \nabla P + \nabla.\overline{\overline \tau} + \vec{ F}_{SF}
\end{equation}

\noindent where $\vec{U }$ is the velocity vector, $  \rho $ is the density, $  \eta $ is dynamic viscosity of fluid. $ p$ denotes pressure. For analysing bubble formation in shear\textendash thinning fluids, a power-law model is considered for estimating the apparent viscosity ($\eta_{app}$), which is expressed as a function of shear rate.\citep{sussman2000} For shear\textendash thinning liquids, the shear stress can be expressed as:
\begin{equation}
\label{eq:tau}
\overline{\overline \tau}= \eta(\dot{\gamma})\dot{\gamma} 
\end{equation}



\noindent where  $\eta$ is a function of all three invariants of the rate\textendash of\textendash deformation tensor. However, in power\textendash law model, the shear\textendash thinning liquid viscosity ($\eta$) is considered to be a function of only shear rate ($\dot{ \gamma }$).
\begin{equation}
\label{eq:nnvis_eqn}
\eta(\dot{\gamma}) =K \dot{\gamma }^{n-1} 
\end{equation} 

\noindent where $K$ and $n$ are the consistency and power-law indices, respectively. The local shear rate ($\dot{ \gamma}$) is related to the second invariant of $\overline{\overline D}$ and is expressed as \cite{fluent}:

\begin{equation}
\label{eq:shearrate}
\dot{ \gamma } = \sqrt{ \frac{1}{2} (\nabla \vec {U} + \nabla { \vec {U} } ^{T})_{ij} (\nabla \vec {U} + \nabla { \vec {U} } ^{T})_{ji}} 
\end{equation} 

\noindent \textbf{Equation of VOF function:} 

\noindent The interface between the gas and the shear\textendash thinning liquid can be traced by solving the following continuity equation of the volume fraction.

\begin{equation}
\label{eq:vof_eqn} 
\frac{\partial    \alpha _{q}  }{\partial t}  +  ( \vec{ U_q }  .  \nabla)   \alpha _{q} =0
\end{equation}

\noindent where $ \alpha _{q}$ is the volume fraction of q phase (gas phase or liquid phase). For a two\textendash phase system, if the phases are represented by the subscripts 1 and 2, and the volume fraction of the phase 2 is known, the density and viscosity in each cell are given by 
\begin{equation}
\rho =  \alpha _{2}  \rho _{2}  + (1 - \alpha _{2})\rho _{1} 
\end{equation}
\vspace{-15px}
\begin{equation}
\eta =  \alpha _{2}   \eta  _{2}  + (1 - \alpha _{2}) \eta  _{1} 
\end{equation}

\noindent The interface between the two phases can be traced by solving the continuity equation for the volume fraction function (Equation \ref{eq:vof_eqn}). The volume fraction of the primary phase can therefore be obtained from the following equation:
\begin{equation}
\label{eq:alpha} 
\sum  \alpha _{q} =1
\end{equation} 

\subsubsection*{Surface tension force}

The continuum surface force (CSF) model\citep{brack-1992} is applied to determine the volumetric surface tension force ($F_{SF})$ term in Equation \ref{eq:mom_eqn}, as follows:

\begin{equation}
\label{eq:source_eqn}
\vec{F} _{SF} = \sigma  \begin{bmatrix} \frac{\mathlarger{\rho } \mathlarger{\kappa_{N}} \mathlarger{\nabla \alpha_o}}{ \mathlarger{\frac{1}{2 }} (\mathlarger{\rho_{o}+  \rho_{w}) }}  \end{bmatrix}    
\end{equation}

\noindent where $\kappa_{N}$ is the radius of curvature and $\sigma$ is the coefficient of surface tension. The interface curvature ($\kappa_{N}$) is calculated in terms of unit normal $\hat{N}$, as:

\begin{equation}
\label{eq:normal_eqn}
\kappa_{N} = - \nabla  .  \hat{N}= \frac{1}{|\vec{ N}|}  \begin{bmatrix} \big( \frac{\vec{ N}}{ |\vec{ N}|} . \nabla\big) |\vec{ N}| - \big( \nabla  . \vec{ N}\big) \end{bmatrix}, \text {where} ~\hat{N} = \frac{\vec{ N}}{ | \vec{ N} | }
\end{equation}

\noindent Moreover, in VOF formulation, surface normal, $N$, is expressed as the gradient of phase volume fraction at the interface, which can be written as: 

\begin{equation}
\centering
\vec{ N}= \nabla  \alpha _{q} 	
\end{equation}

\noindent This surface tension force is employed by the piecewise-linear interface calculation (PLIC) scheme, which accurately computes curvatures for reconstruction of the interface front.\cite{fluent,guo2015} Wall adhesion effect is considered by defining a three-phase contact angle at the channel wall ($\theta_{W}$). Consequently, the surface normal at the reference cell next to the wall is calculated as:

\begin{equation}
\label{eq:normal1_eqn}
\hat{N}=  \hat{N}_{W}  cos \theta_{W}  +  \hat{M}_{W} sin \theta _{W} 
\end{equation}

\noindent where $\hat{N}_W$ and $\hat{M}_W$ are the unit vectors normal and tangential to the wall, respectively.\cite{fluent} In VOF, it is difficult to capture the geometric properties (interface normal and curvature) from the VOF function whose spatial derivatives are not continuous near the interface. Such inaccurate calculations of geometric properties may lead to spurious currents. Therefore, the volumetric surface tension force ($F_{SF})$ term is modified with a continuous LS function to reduce spurious currents that helps in improving radius of curvature estimation, as mentioned in the subsequent section.

\subsection*{Equation of level set (LS) function:} 
\begin{equation}
\frac{\partial   \varphi }{\partial t}  +  	\vec{ U_q } .  \nabla  \alpha _{q}=0  
\end{equation}

\noindent where,  $\varphi $ is the level function \citep{suss-1994}, $\alpha$ is the smoothed volume fraction in the $q_{th}$ cell, the smoothed volume fraction field in the CLSVOF method is defined using a smoothed Heaviside function ($\textit{H}(\varphi)$) \citep{sussman2000}, defined as

\begin{equation}
\textit{H}(\varphi)=\begin{cases} 0& \textit{ if} \hspace{5px} \varphi  < - a  \\  \frac{1}{2}[1+ \frac{ \varphi }{a} + \frac{1}{ \pi }sin( \frac{ \pi  \varphi }{a} ) ]  & \textit{ if} \hspace{5px} |  \varphi  |  \leq a\\ 
1 & \textit{ if} \hspace{5px} \varphi  > a
\end{cases}  
\end{equation}
\noindent The level function ($\varphi$) is a function of position vector ($\vec{\chi}$) and time ($t$), and \textit{a} is the interface thickness. The level set function $\varphi ( \vec{\chi },{{t} } )$ acts as the signed distance from the interface \citep{suss-1994}.   

\begin{equation}
\varphi ( \vec{\chi },{{t} } ) = \begin{cases}d  & \text{if} \hspace{5px} \text{$\chi$ is  in the liquid phase}\\0 & \text{if} \hspace{5px} \text{$\chi$ is  in the interface}\\-d & \text{if} \hspace{5px} \text{$\chi$ is  in the gas phase} \end{cases}  
\end{equation}

\noindent where $\textit{d}$ is the shortest distance of a point $\vec{\chi}$ from interface at time $\textit{t}$. 

\noindent The fluid type is identified based on the sign of the level set function. It takes positive values in the liquid region, negative values in the gas region and zero value at the interface. To solve the Navier\textendash Stokes equation, the fluid properties density and viscosity distribution in the whole solution domain are required. Since the density of each fluid is constant, it takes two different values depending on the sign of level set function. The physical properties of the mixture are calculated using a smoothed Heaviside function $\textit{H}(\varphi)$ \citep{sussman2000}, so that the properties vary continuously across the interface, as follows:

\begin{equation}
\rho ( \varphi )=  \textit{H}( \varphi ) \rho _{2}  + (  1-\textit{H}( \varphi ) )   \rho _{1}  
\end{equation} 
\begin{equation}
\mu ( \varphi )=   \textit{H}( \varphi ) \mu  _{2}  + (   1 -  \textit{H}( \varphi )  ) \mu  _{1}
\end{equation}

\noindent In the CLSVOF method, the normal vector  ($\hat{N}$) in  radius of  curvature term (Equation\ref{eq:normal_eqn}) is computed by level set function ($\vec{ N}= \nabla  \varphi$). The main advantage of using the LS function is its ability to determine an accurate unit normal vector to the interface \citep{sussman1999}  ${\hat{N} = \frac{\mathlarger \nabla \mathlarger \varphi}{ | \mathlarger \nabla  \mathlarger \varphi | }} $, and hence the interface curvature can be calculated precisely as:

\begin{equation}
\kappa ( \varphi )= \nabla .  \frac{ \nabla  \varphi }{ |  \nabla  \varphi  | } 
\end{equation}
\begin{equation}
\delta ( \varphi )=\begin{cases}0 & \textit{ if} \hspace{5px} |  \varphi  |  \geq a\\ \frac{1}{2a} (1+cos( \frac{ \pi  \varphi }{a} )) & \textit{ if} \hspace{5px} |  \varphi  |  < a\end{cases} 
\end{equation}

\noindent Finally, the volumetric surface tension force is estimated using the level set function. Therefore, the volumetric surface tension ($\vec{ F}_{SF} $) in Equation \ref{eq:mom_eqn} based on CSF method is calculated as: \citep{sussman1999} 

\begin{equation}
\vec{ F}_{SF} = \sigma  \kappa ( \varphi ) \delta ( \varphi ) \nabla \varphi 
\end{equation}

\noindent where $\sigma $ is the surface tension, $\kappa ( \varphi )$ is the interface curvature, $\delta ( \varphi )$ is the Dirac Delta function. 

\subsection*{Comparison of VOF and CLSVOF methods}

To understand the benefit of CLSVOF over the VOF, in this section, the order of accuracy in both methods are compared for a gas-liquid (air-water) system. A three dimensional stationary bubble rise in the liquid system is considered for this test case. A bubble of radius $R=5~mm$ is initially placed at the center of a 4R$\times$4R$\times$4R cubic box. Typical air-water system fluid properties are considered. A constant pressure boundary condition is applied on all the four boundaries and time step size is taken as $10^{-8}$ s for the simulation.     
\begin{figure}[!ht]
	\centering
	\includegraphics[width=\textwidth]{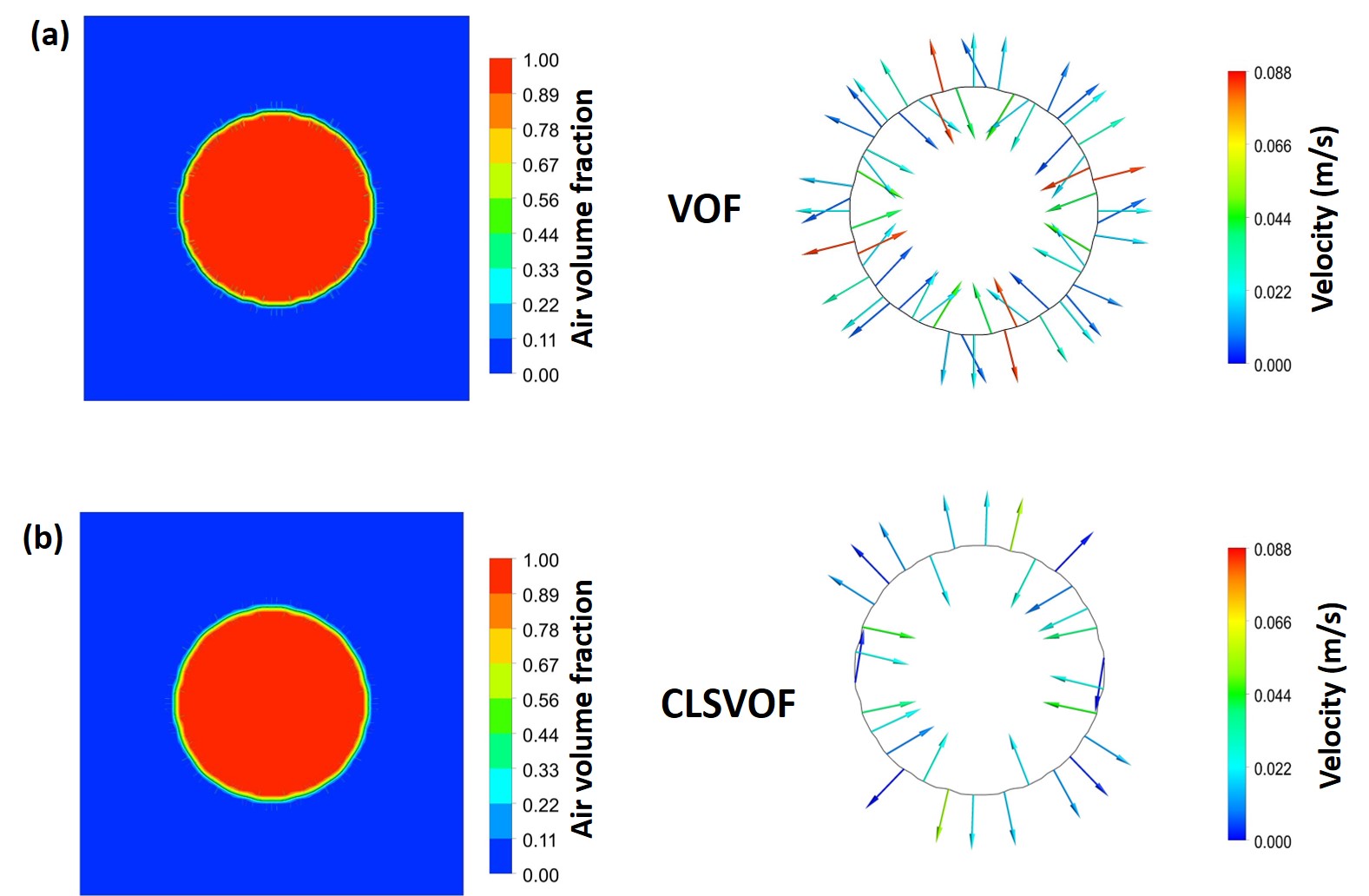}
	\caption{\label{fig:A1}Comparison of parasitic currents around the interface resulting from (a) VOF method and (b) CLSVOF method. (red color: gas phase, blue color: liquid phase).}
\end{figure}
Figure \ref{fig:A1} shows the comparison of interface and velocity fluctuations around the interface for VOF and CLSVOF methods. It is evident from Figure \ref{fig:A1}a that the VOF simulation captures the wavy interface. This can mainly be attributed to the imbalance between the surface tension force and sudden pressure rises across the interface resulting parasitic currents. However, in CLSVOF method with identical number of mesh elements, a smooth interface is captured and the magnitude of velocity fluctuations is found to significantly decrease, as illustrated in Figure \ref{fig:A1}b. It shows that the implementation of CLSVOF method can lead to minimization of the spurious currents around the interface, which can be significant in low Capillary number problems. Dang et al.,\citep{dang-2015} also reported the accuracy of CLSVOF over VOF in terms of interface tracking which were substantiated with their experimental visualization. Therefore, in the present study, CLSVOF method is used to understand the slug-flow formation in shear-thinning liquids.	

\subsection*{Computational model} 
\begin{figure}[h]
	\centering
	\includegraphics[width=0.6\textwidth]{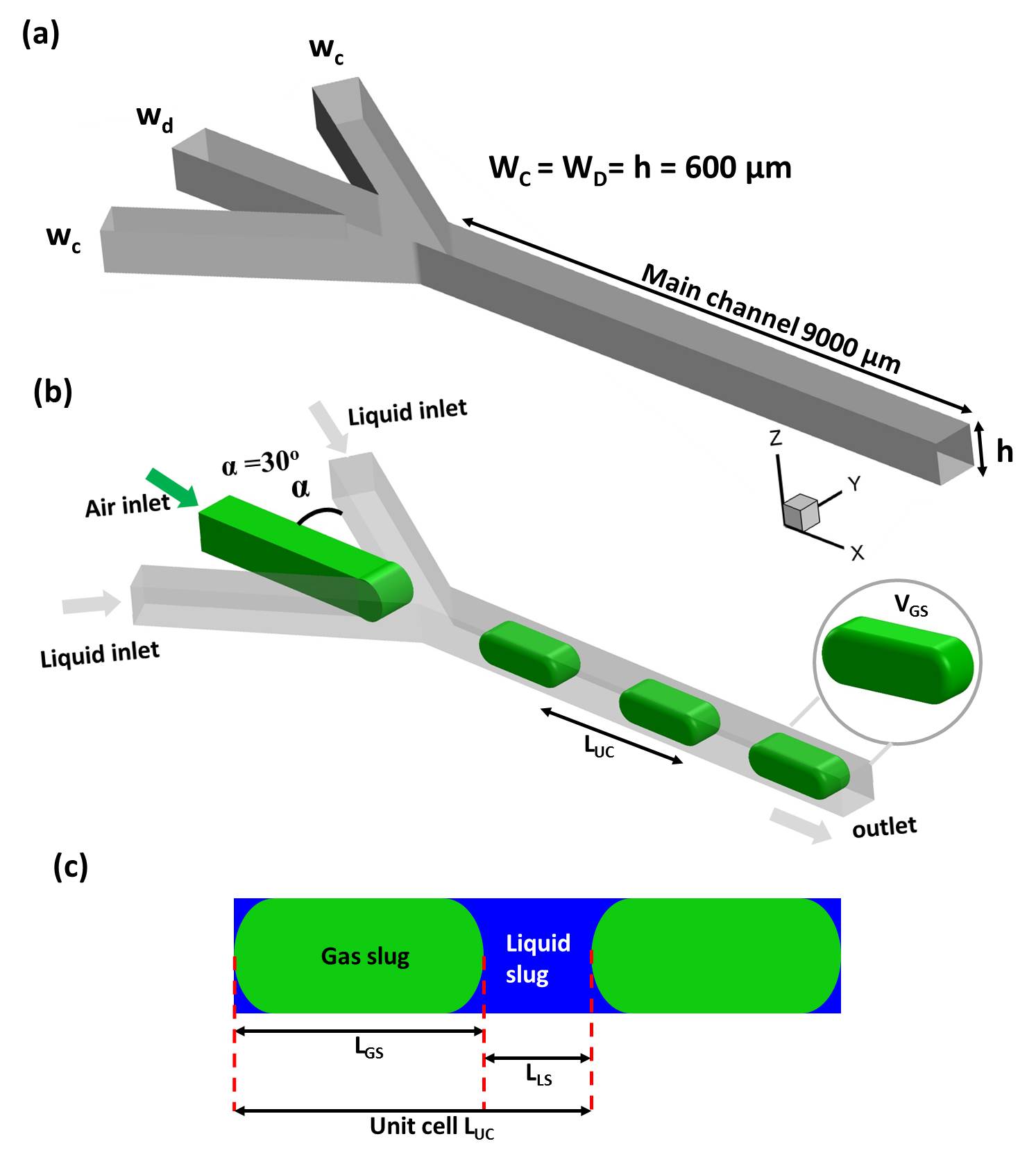}
	\caption{\label{fig:M1}Schematic of a (a) 3D converging microchannel, (b) 3D representation of slug flow, and (c) unit cell model in segmented flow, where $L_{GS}$, and $L_{LS}$ are lengths of gas and liquid slugs, respectively.}
\end{figure}

Figure \ref{fig:M1}a shows the schematic of a 3D converging microchannel that is considered in this study to investigate the bubble formation in shear\textendash thinning liquid. The main channel length is considered as 9000 $\mu$m with a uniform cross-section of 600 $\mu$m\texttimes 600 $\mu$m. This configuration and dimensions are adopted from the experimental work of Dang et al. \cite{dang-2015,dang-2013} 
The gas phase flows through the middle channel ($w_d$), and the continuous phase shear\textendash thinning liquid is introduced through two side inlets having an identical cross\textendash section of 600 $\mu$m\texttimes 600 $\mu$m, as well. Gas and liquid inlets are 3 mm long, and the angle between two inlets is 30\textdegree. At the merging junction, shear force of the continuous phase (shear\textendash thinning liquid) acts on the dispersed phase (gas) that results in gas slug formation, which flows through the main channel, as illustrated in Figure \ref{fig:M1}b. Subsequently, the segmented flow phenomena is analysed based on gas slug length ($L_{GS}$), gas slug volume ($V_{GS}$), and unit cell length ($L_{UC}$), as distinguished in Figure \ref{fig:M1}b and c. The time-dependent governing equations are solved in a finite volume method based CFD solver, Ansys Fluent 17.0.\cite{fluent} Pressure implicit with splitting operators (PISO) logarithm is used to solve the pressure\textendash velocity coupling in momentum equation.\citep{issa1986} The spatial derivatives in momentum and level set equations are discretized using the second-order upwind scheme.\citep{barth1989} Volume fraction is solved using piecewise linear interface construction (PLIC) geometrical reconstruction algorithm.\citep{holt2012} Variable time step and fixed Courant number (Co = 0.25) are considered for solving the governing equations with constant velocity boundary condition for both gas and liquid inlets. At the flow exit, pressure outlet boundary condition is specified. It is assumed that the continuous phase completely wets the channel wall and the slug formation occurs at the junction. Consequently, the solid walls are set to no\textendash slip boundary condition with a static contact angle specified throughout the wall which is in line with other recent numerical studies \cite{kagawa2014permeation,zhang2014effects}.

\section*{Results and discussion}


At first, grid independence study is carried out using three different number of mesh elements for dimensionless gas slug length ($L_{GS}/D_H$) estimation. Table \ref{tbl:grid} indicates 70,000 cells as the optimum number of mesh elements for the considered geometry. It is worth mentioning that, in our previous works \cite{sontti2017cfd,sontti2017numerical,sontti2018formation} we captured liquid film thickness precisely around the Taylor bubble. We implemented similar modeling strategy in this work and attempted to capture liquid film thickness for the square microchannel with systematic wall refinement. However, in the range of flow conditions studied in this work, the liquid film thickness was not captured even with extremely fine meshes. Therefore, in the present study, we mainly focus our investigations towards non\textendash Newtonian flow behavior in a converging microchannel to realize the effect of rheological properties on the slug length, volume and flow profiles, thereby neglecting liquid film thickness (assuming the thickness of the liquid film between the microchannel wall and the plug surface is negligible compared to the microchannel width and depth \cite{sattari2017hydrodynamics,raimondi2014experiments}).

\begin{table} [h]
	\centering
	\noindent \caption{ Effect of mesh size on dimensionless gas slug length at $J_G$/ $J_L$= 1, $J_{TP}$= 0.217 m/s, $\eta_L$= 0.00983 Pa.s, $\sigma$= 0.0726 N/m, and $\theta$= $60$\textdegree.}
	\label{tbl:grid}
	\begin{tabular*}{0.55\textwidth}{@{\extracolsep{\fill}}lll}
		
		\hline
		S.no &Mesh elements  &  $L_{GS}/D_H$ \\ 
		\hline
		1 & 50,000 & 3.233    \\
		2 & 70,000 & 3.195    \\
		3 & 90,000 & 3.195  \\
		\hline
	\end{tabular*}
\end{table} 

\begin{figure}[h]
	\centering
	\includegraphics[width=0.75\textwidth]{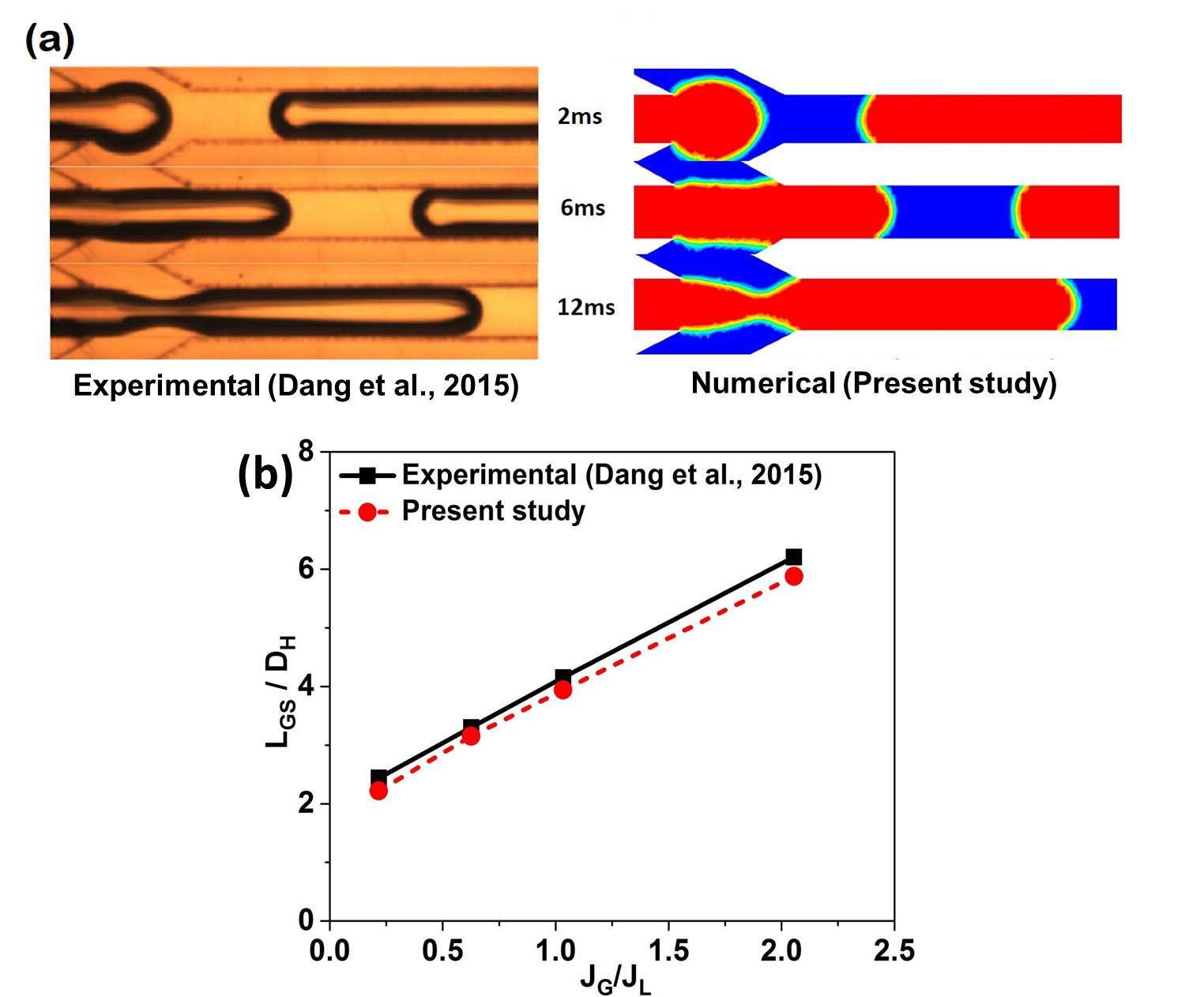}
	\caption{\label{fig:VL} Comparison of (a) gas slug evolution with flow time, and (b) non-dimensional gas slug length with the experimental data.\citep{dang-2015}. (red color: gas phase, blue color: liquid phase).}
\end{figure}

\noindent The developed model is initially validated with the experimental results of Dang et al., \cite{dang-2015} for a Newtonian system. The fluid properties considered in model validation are listed in Table \ref{tbl:Dang_data}. 
\begin{table} [h]
	\centering
	\noindent \caption{Fluid properties used in model validation.\cite{dang-2015}}
	\label{tbl:Dang_data}
	\begin{tabular*}{0.88\textwidth}{@{\extracolsep{\fill}}llll}
		\hline
		Fluid  & Density $(kg/m^{3})$ &Viscosity (Pa.s) & Surface tension (N/m) \\ 
		\hline
		Air & 1.225 & 1.7894 $\times 10^{-5}$ & \textendash   \\
		Water & 1000 & 1 $\times 10^{-3}$ &  0.0726 \\
		\hline
	\end{tabular*}
\end{table}
Figure \ref{fig:VL} portrays the comparison of model prediction and experimental data of gas slug length, and the maximum deviation is observed to be less than 9.0\%. It is noteworthy that experimental studies on the slug flow formation in non-Newtonian liquids flowing in a converging shaped configuration have not been reported in the literature. Due to the lack of experimental data, several researchers have demonstrated Newtonian model validation for prediction of non-Newtonian flow behavior \cite{chen-2013, islam2015pair, madadelahi2018newtonian}. Accordingly, our validated model with Newtonian systems has been extended here to understand bubble formation in non-Newtonian liquid.

\noindent The validated model is thereafter extended for the shear\textendash thinning system. To examine the effect of rheology in segmented flow, a two-parameter power\textendash law model has been considered. Aqueous solutions of Polyacrylamide (PAM) with different mass concentrations (0.1, 0.2 and 0.4 wt\%) are selected as the shear\textendash thinning liquids, which exhibit power\textendash law behaviour. Rheological properties of those solutions are adopted from the experimental measurements of Mansour et al., \cite{manso-2015} and are listed in Table~\ref{tab:power_law_data_picchi}. It can be observed from Table  \ref{tab:power_law_data_picchi}, that differences in density and surface tension values in those solutions are negligible. Therefore, such selection of liquid phases will help to understand the influences of either power\textendash law index ($n$), or consistency index ($K$), when varied separately.

\begin{table}[h]
	\centering
	\noindent \caption{Rheological properties of PAM with different mass concentrations.\citep{manso-2015}} 
	\label{tab:power_law_data_picchi}
	\begin{tabular}{@{}lllll@{}}
		\hline
		\begin{tabular}[c]{@{}l@{}}Liquid phase\\ (PAM\textendash wt\% Conc.)\end{tabular} & \begin{tabular}[c]{@{}l@{}}Density, \\ $\rho$ ($kg/m^{3}$)\end{tabular} & \begin{tabular}[c]{@{}l@{}}Power\textendash law\\ index, $n$  (\textendash)\end{tabular} & \begin{tabular}[c]{@{}l@{}}Consistency \\ index, $K$ ($Pa.s^{n}$)\end{tabular} & \begin{tabular}[c]{@{}l@{}}Surface tension,\\  $\sigma$ (N/m) \end{tabular} \\ \hline
		PAM\textendash0.1\%   & 998 & 0.99& 0.00176 & 0.072 \\
		PAM\textendash0.2\%  & 999 & 0.94 & 0.00365  & 0.073 \\
		PAM\textendash0.4\%  & 999 & 0.85 & 0.014 & 0.073 \\
		
		\hline
		
	\end{tabular}
\end{table}

\subsection*{Effect of PAM concentration}

\begin{figure}[h]
	\centering
	\includegraphics[width=1.0\textwidth]{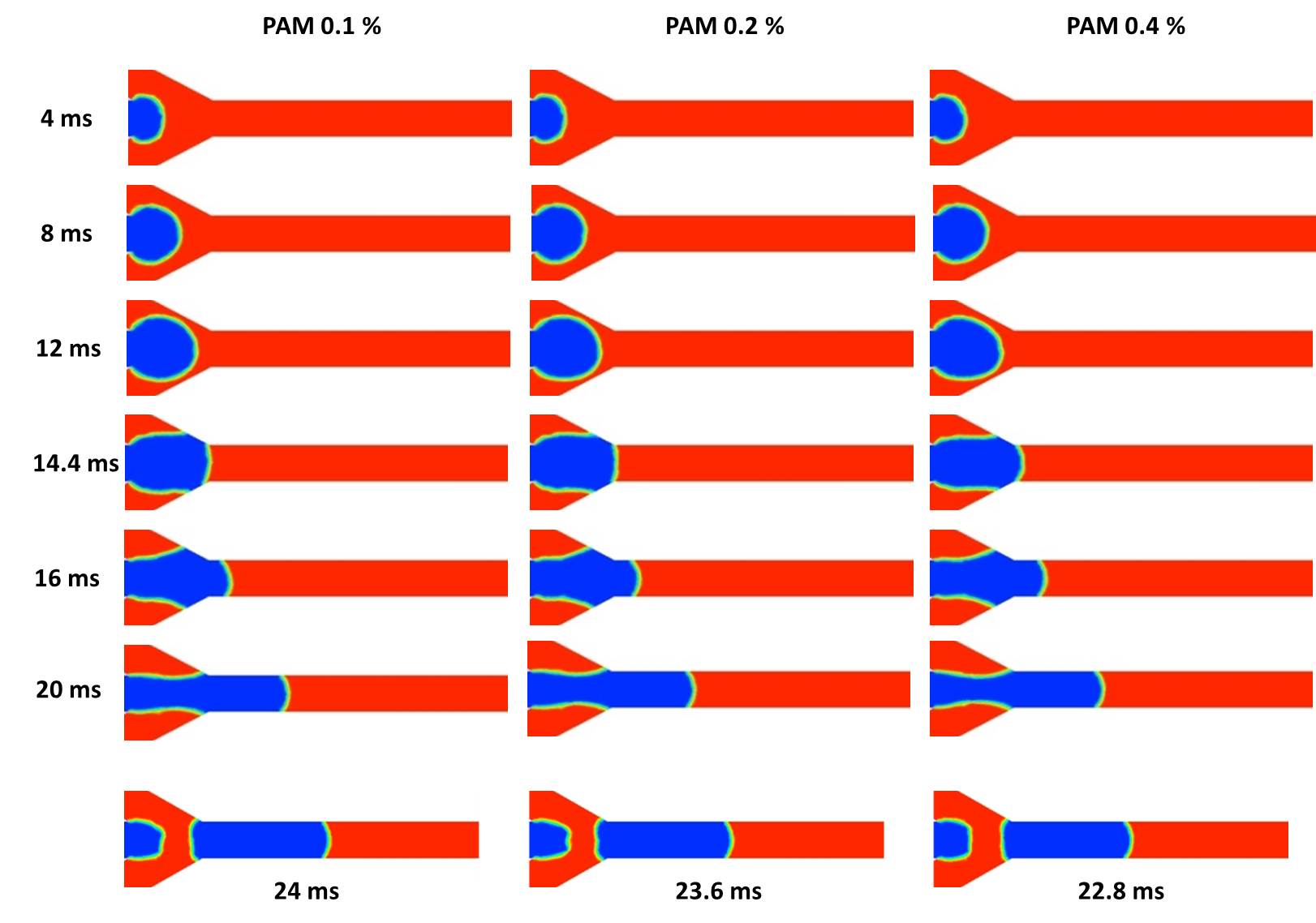}
	\caption{\label{fig:PAM1} Formation of segmented flow for three concentrations of PAM solutions at $J_G$= 0.1085 m/s, $J_L$= 0.1085 m/s, and $\theta$= 60\textdegree. (red color: liquid phase, blue color: gas phase).}.
\end{figure}

Figure \ref{fig:PAM1} illustrates the formation of segmented flow for three different concentrations of PAM solution. Typically, gas slug formation process consists of a two\textendash step mechanism. Firstly, the emerging gas slug expands axially and radially until its tip blocks the entrance of the main channel, known as the expansion step. Subsequently, the axial velocity component of the liquid from two side inlets of the microchannel along with the gas dynamic pressure, drive the emerging gas slug into the main microchannel under the action of shear stress. The pressure difference across the gas\textendash liquid interface squeezes the emerging gas slug to form a neck at the mixing junction until the neck eventually ruptures, which is known as the rupture step.
\noindent In this described mechanism, various forces are exerted on emerging gas slug such as surface tension, viscous and inertial forces.\citep{dang-2013} Figure \ref{fig:PAM1} also depicts that as PAM concentration increases gas slug formation time reduces. This can be attributed to the increased effective or apparent viscosity of the solution (as estimated by Equation \ref{eq:effective_eqn}) increases with increasing PAM concentration, and intensifies the shear stress. 
\begin{equation}
	\label{eq:effective_eqn}
	\eta_{eff}=K\left(\frac{3n+1}{4n}\right)^{n}\left(\frac{8J_{L}}{W_c}\right)^{n-1}
\end{equation}
\noindent where $K$ is consistency index, $J_L$ is liquid  velocity, $W_c$ is diameter of the channel, and $n$ is power\textendash law index.

\begin{figure}[h]
	\centering
	\includegraphics[width=0.65\textwidth]{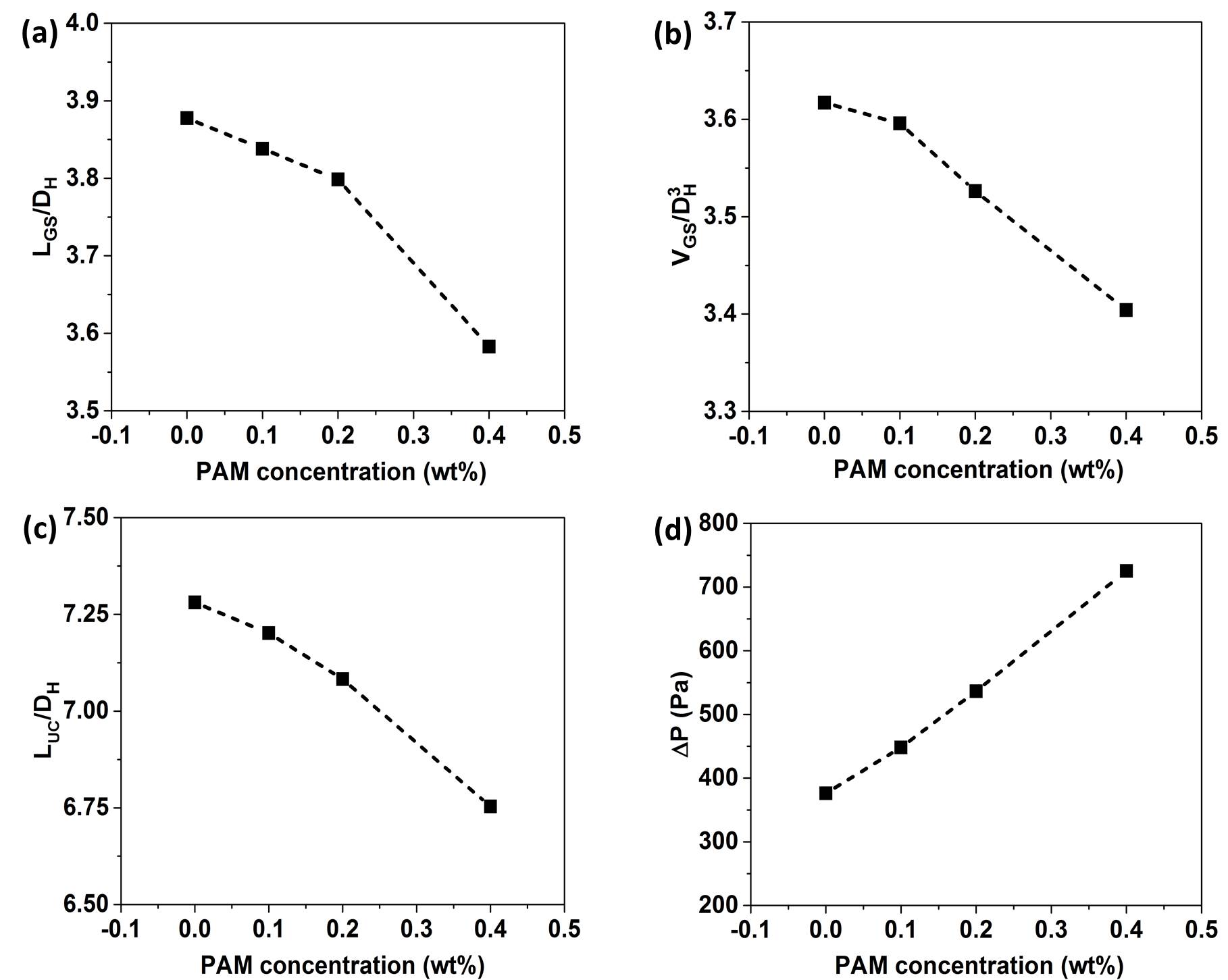}
	\caption{\label{fig:PAM2} Effect of PAM concentration on (a) gas slug length, (b) gas slug volume, (c) unit cell length, and (d) pressure drop at $J_G$= 0.1085 m/s, $J_L$= 0.1085 m/s, and $\theta$= 60\textdegree.}
\end{figure}

\noindent Furthermore, at the end of expansion step, the front of emerging gas slug appears to block the main channel completely, but due to square cross\textendash section of the microchannel, the liquid can still flow through the gap between the emerging slug and the channel wall, as shown in Figure \ref{fig:PAM1}. Velocity of liquid flowing through this gap is expected to be larger than that of the emerging gas slug. Consequently, significant shear stress caused due to relatively large velocity difference at the interface would act on the emerging gas slug that speeds up the rupture process, which is especially the case at higher liquid viscosity.\citep{van-2007} 
Figure~\ref{fig:PAM2} quantitatively depicts the effect of PAM concentration on gas slug length, volume, unit cell length, and pressure drop. 
It is evident from Figure \ref{fig:PAM2}a\textendash b that gas slug length and volume decrease with increasing PAM concentration due to enhanced effective viscosity, as discussed earlier. In other microchannel configurations, similar observation was also experimentally and numerically reported by several researchers. \cite{fu2011gas,manso-2015,sontti2018formation} As shape of the gas slug is not altering, Figure \ref{fig:PAM2}c reveals similar trend for unit cell length, which is the combination of a gas slug and a liquid slug length (see Figure \ref{fig:M1}c). The pressure drop, measured from inlet of gas phase to the outlet of the main channel, increases with PAM concentration, as depicted in Figure \ref{fig:PAM2}d. This can be ascribed to higher frictional pressure drop exerted by the enhanced viscous nature of higher concentration solutions. 

\begin{figure}[!ht]
	\centering
	\includegraphics[width=0.9\textwidth]{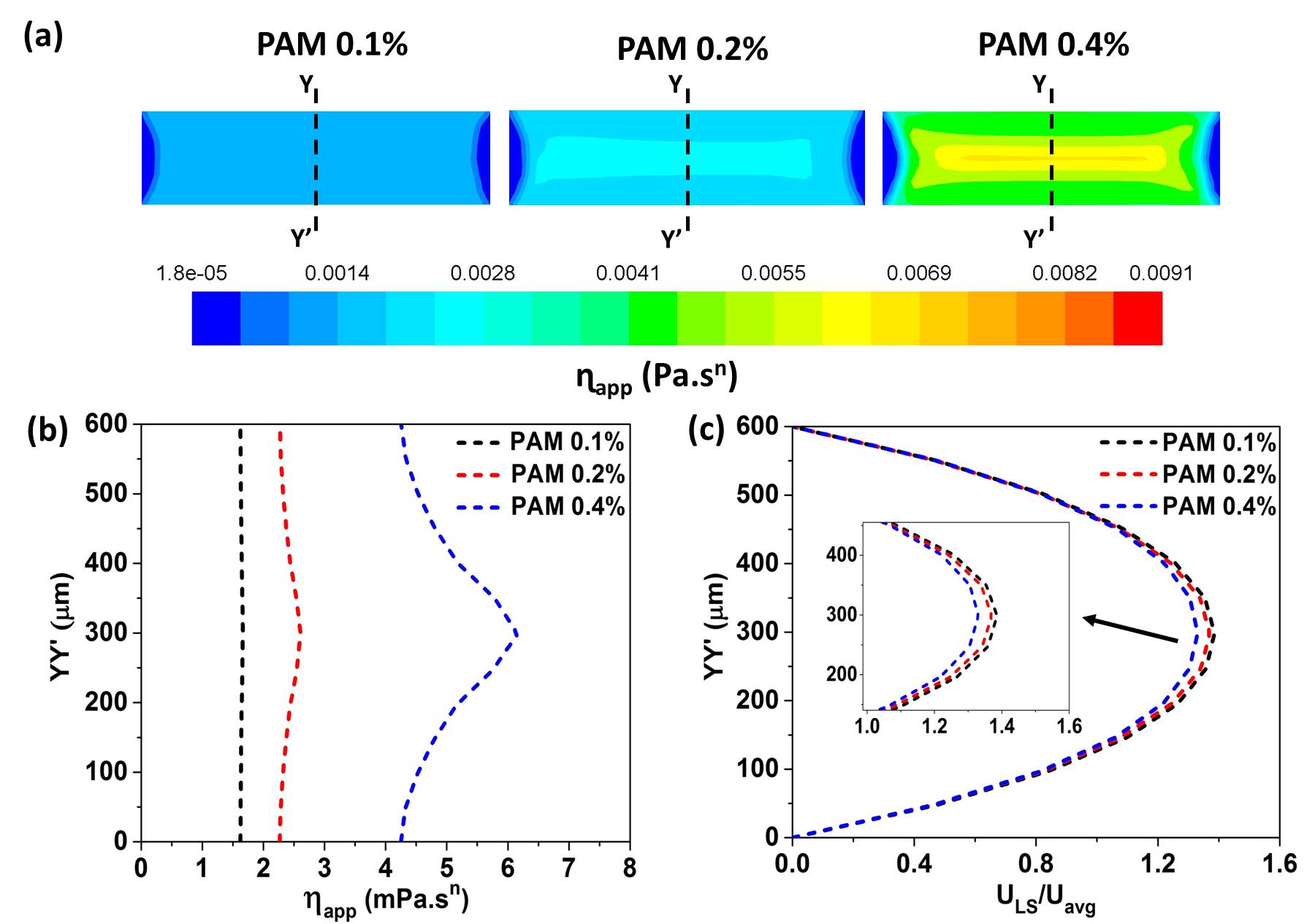}
	\caption{\label{fig:PAM3} (a) Contours of apparent viscosity distribution inside liquid slug, (b) apparent viscosity, and (c) velocity profile on YY' plane inside liquid slug at $J_G$= 0.1085 m/s, $J_L$= 0.1085 m/s, and $\theta$= 60\textdegree.}
\end{figure}

\noindent Figure \ref{fig:PAM3}a demonstrates contours of apparent viscosity variation inside the liquid slug that decreases with increasing shear rate, and shows a maximum value at the centre of the channel due to lower shear rate. 
Analysis of apparent viscosity over YY'plane inside liquid slug (marked by dotted lines in Figure \ref{fig:PAM3}a) shows that for PAM 0.1 wt\% apparent viscosity change is negligible. However, as concentration increases, noticeable variation in viscosity profile is observed. In the case of PAM 0.1\% closer to parabolic profile is evident in the slug. However, a relatively flatter profile (i.e., with lower magnitude) is evident for PAM 0.2\% and PAM 0.4\% solutions. These are the typical characteristic of the laminar velocity profiles of shear-thinning fluids. Subsequently, velocity profiles inside the liquid slug are also analysed on YY' plane. From Figure \ref{fig:PAM3}c it is observed that with increasing concentration velocity profile at the centre of the channel becomes flatter. Shear\textendash thinning fluids exhibit such profile in the middle of the channel under fully developed condition, which is affirmed in Figure \ref{fig:PAM3}c. The change in velocity profiles is not substantial here, as it can be due to weaker shear thinning nature of PAM solutions. 
\subsection*{Effect of velocity ratio}
\begin{figure}[h]
	\centering
	\includegraphics[width=0.8\textwidth]{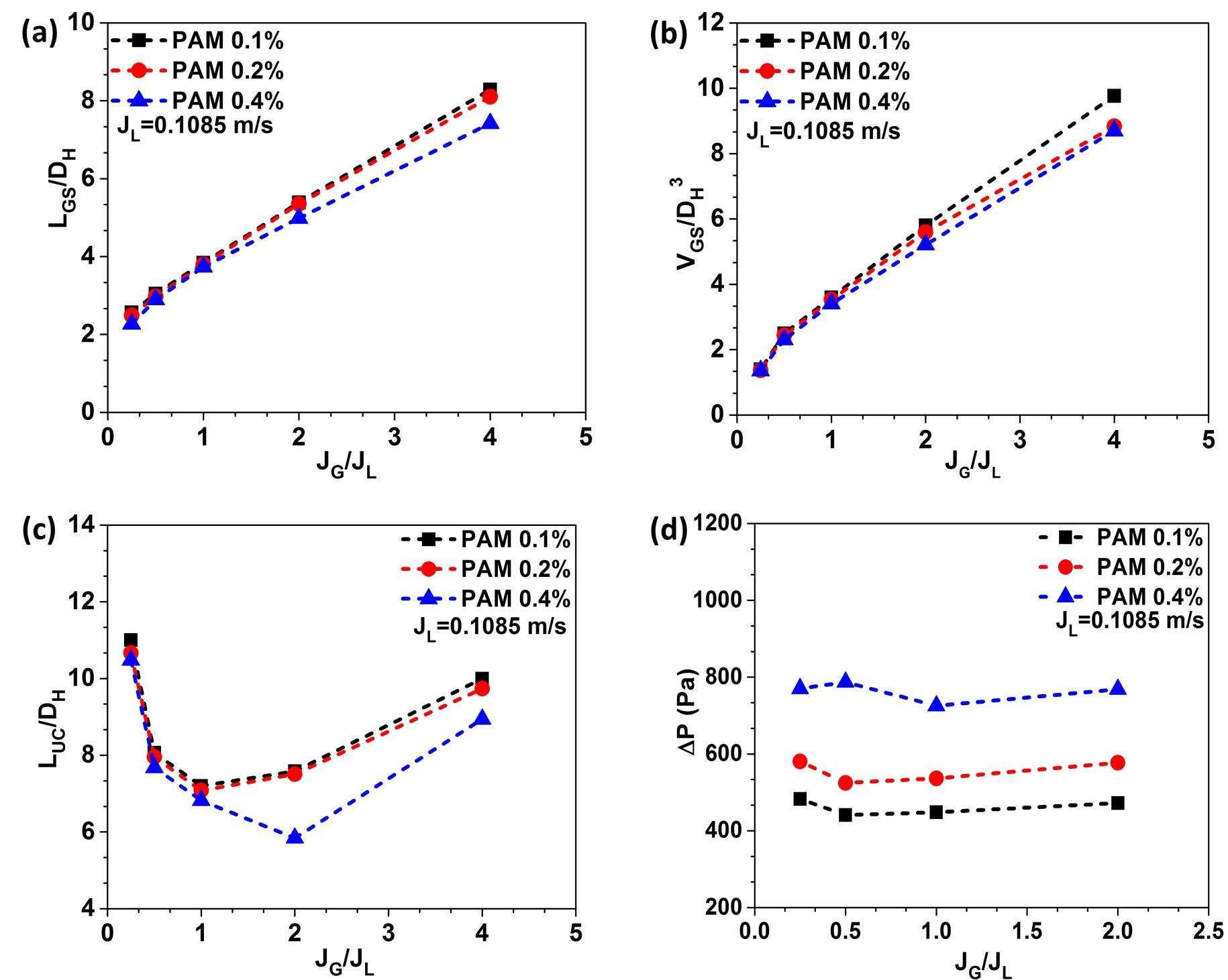}
	\caption{\label{fig:liquid} Effect of velocity ratio at fixed liquid velocity $J_L$= 0.1085 m/s on (a) gas slug length, (b) gas slug volume, (c) unit cell length, and (d) pressure drop for $\theta$= 60\textdegree. }
\end{figure}

\begin{figure} [!ht]
	\centering
	\includegraphics[width=\linewidth]{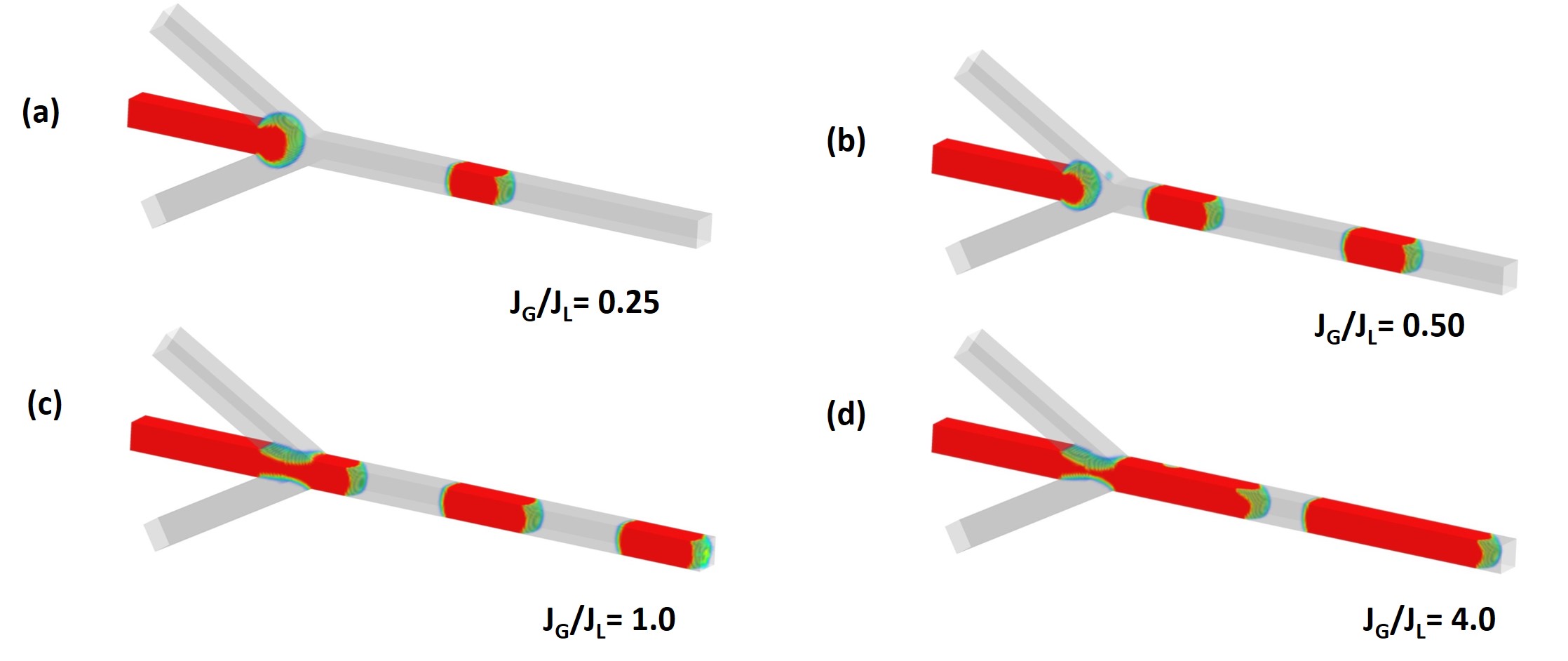}
	\caption{\label{fig:gasveleffect} Effect of gas phase inlet velocity on slug formation for PAM 0.4 wt\% (a) $J_G$/$J_L$= 0.25, (b) $J_G$/$J_L$= 0.50, (c) $J_G$/$J_L$= 1.0, and (d) $J_G$/$J_L$= 4.0 at fixed $J_L$= 0.1085 m/s, and $\theta$= 60\textdegree. (red color: gas phase).}
	
\end{figure}

The effect of gas and liquid velocities are analysed by keeping other fluid properties constant. Figure \ref{fig:liquid}a shows the effect of gas velocity on the gas slug length at a fixed liquid velocity ($J_L$=0.1085 m/s).
For all concentrations as the $J_G$/$J_L$ ratio increases, the gas slug length and volume increase due to enhanced gas phase velocity. However, there are negligible differences in gas slug length and volume at lower velocity ratios ($J_G$/$J_L$) among the PAM solutions (Figure~\ref{fig:liquid}b). Interestingly, significant variation in gas slug length and volume are observed at higher $J_G$/$J_L$ owing to the dominance of higher inertial forces of the gas phase over the viscous force. Figure \ref{fig:liquid}c shows that for all the PAM concentrations, unit cell length attains a minimum value as the gas slug length transition occurs from very small to quite larger values at a higher $J_G$/$J_L$. Due to the viscous nature of PAM 0.4 wt\% solution, this minimum (of $L_{UC}/D_H$) is observed at $J_G$/$J_L$=2, while for low concentration solutions, it is observed at $J_G$/$J_L$= 1. It is observed that the pressure drop is least influenced by increasing $J_G$ at a fixed $J_L$, as shown in Figure \ref{fig:liquid}d. Figure \ref{fig:gasveleffect} qualitatively describes that with increasing the gas phase velocity gas slug length increases, as mentioned in Figure~\ref{fig:liquid}a.

\begin{figure}[!ht]
	\centering
	\includegraphics[width=0.8\textwidth]{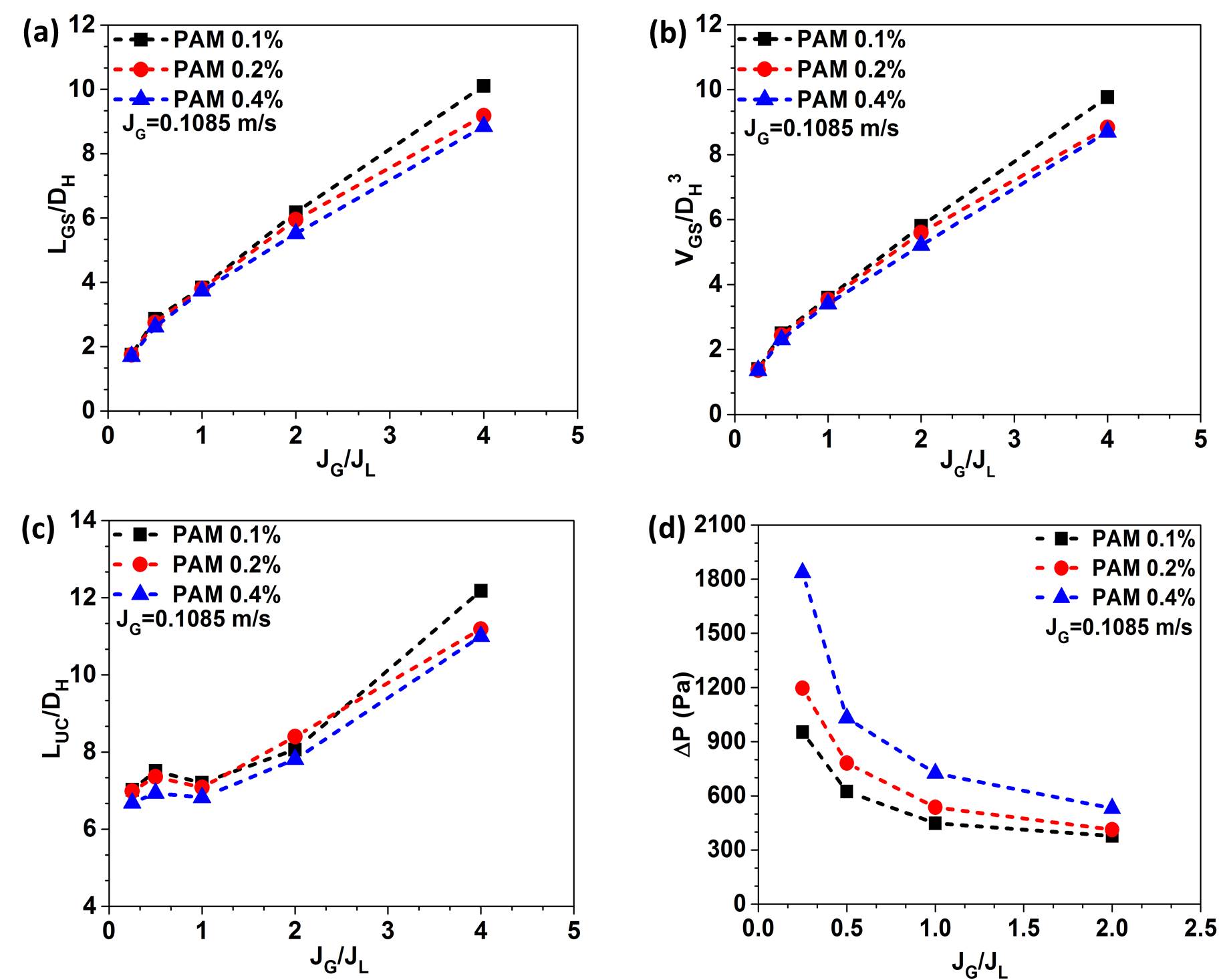}
	\caption{\label{fig:gas} Effect of velocity ratio at fixed gas velocity $J_G$= 0.1085 m/s on (a) gas slug length, (b) gas slug volume, (c) unit cell length, and (d) pressure drop at $\theta$= 60\textdegree.}
\end{figure}
\begin{figure} [!ht]
	\centering
	\includegraphics[width=\linewidth]{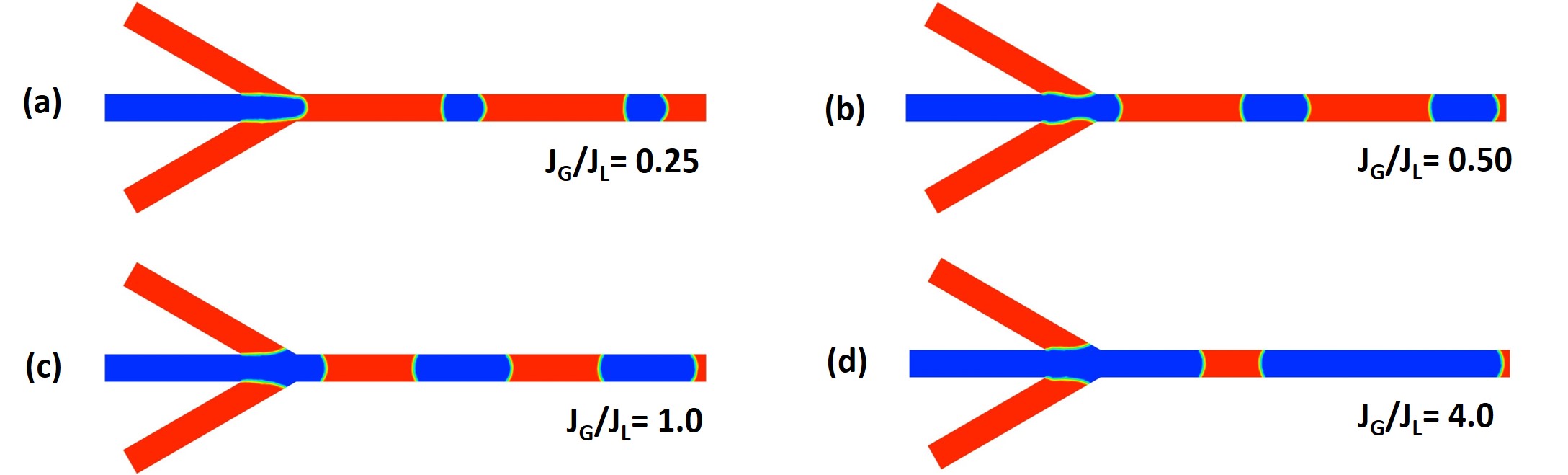}
	\caption{\label{fig:liqveleffect} Effect of liquid phase inlet velocity on slug formation for PAM 0.4 wt\% (a) $J_G$/$J_L$= 0.25, (b) $J_G$/$J_L$= 0.50, (c) $J_G$/$J_L$= 1.0, and (d) $J_G$/$J_L$= 4.0 at $J_G$= 0.1085 m/s, and $\theta$= 60\textdegree. (red color: liquid phase, blue color: gas phase). }
	
\end{figure}

 
\noindent Effect of liquid phase velocity ($J_L$) is also studied by keeping the gas phase velocity ($J_G$) and other fluid properties constant. Figure \ref{fig:gas}c depicts that as the velocity ratio is increased for a fixed $J_G$, unit cell length for all concentrations increases while the range it covers is comparatively lesser than the gas slug length (refer to Figure \ref{fig:gas}a) for similar velocity ratios. This trend confirms that the liquid slug length decreases with increase in $J_G$/$J_L$.
Figure \ref{fig:gas}d shows reduction in pressure drop values for all concentrations with increasing $J_G$/$J_L$ at a fixed $J_G$, unlike nearly constant pressure drop values obtained by varying $J_G$/$J_L$ at a fixed $J_L$, as discussed in the previous section. This indicates that the global pressure drop is mostly influenced by the continuous liquid, which is striking for higher concentration solutions due to their intrinsic higher effective viscosity. Moreover, the pressure drop is substantial at lower velocity ratios or larger liquid velocities. Figure \ref{fig:liqveleffect} shows the volume fraction contours influenced by the liquid velocity, where the gas slug length increases with decreasing liquid phase velocity at a constant gas flow rate.

\subsection*{Effect of surface tension}
Segmented flow in microchannels has a strong dependence on the surface tension and viscous forces. To understand its effect, in this study the surface tension is varied from 0.030 to 0.090 N/m for all PAM solutions, which are presented in the form of respective modified Capillary number ($Ca^{*}$=$KU_B^n{D_H}^{(1-n)}/\sigma$). Generally, with changing surface tension, contact angle may alter significantly. However, to realize the sole effect of the surface tension, the contact angle was unaltered. 
\begin{figure}[!ht]
	\centering
	\includegraphics[width=0.8\textwidth]{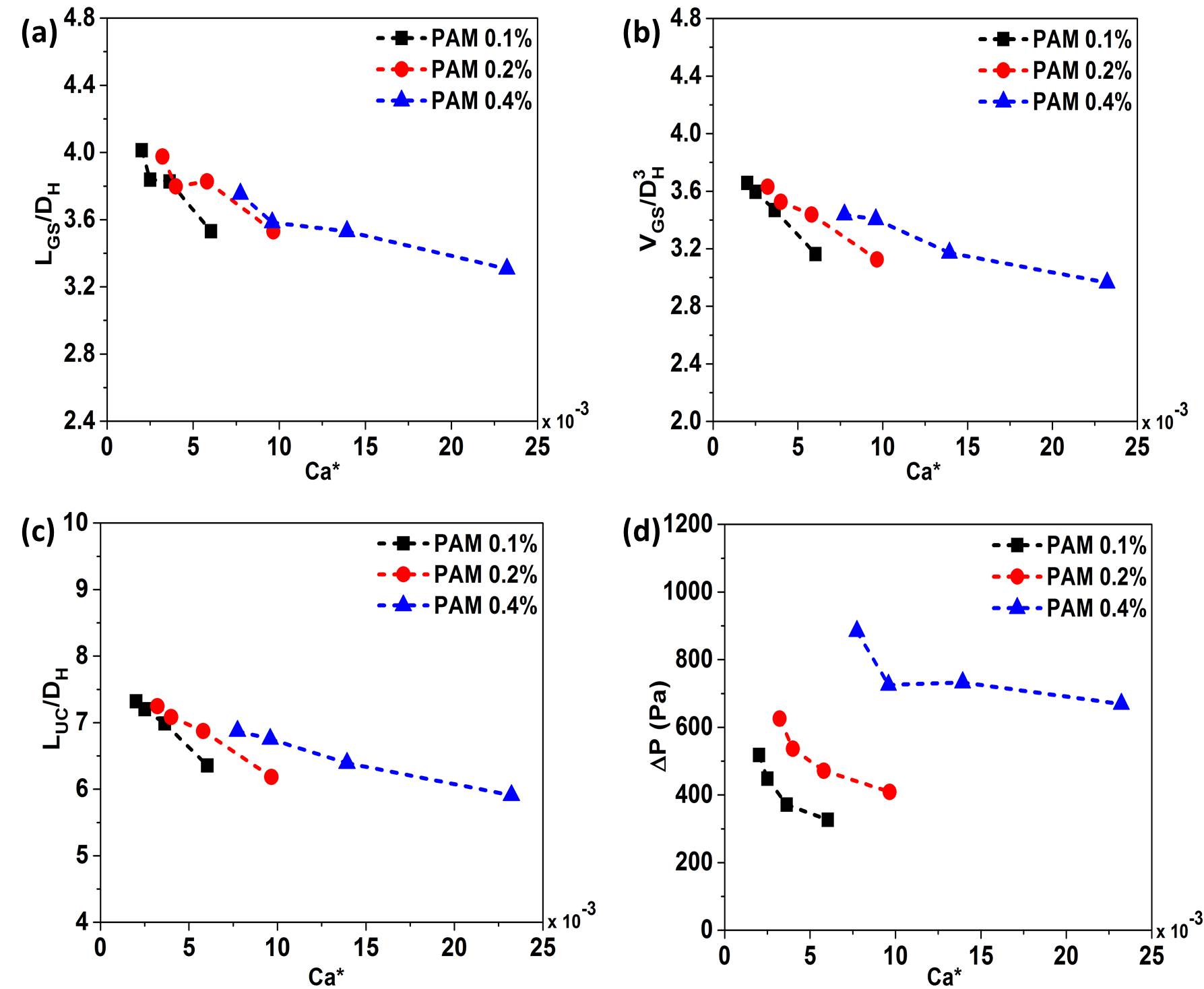}
	\caption{\label{fig:ST1} Effect of surface tension ($\sigma$) on (a) gas slug length, (b) gas slug volume, (c) unit cell length, and (d) pressure drop at $J_G$= 0.1085 m/s, $J_L$= 0.1085 m/s, and $\theta$= 60\textdegree.}
\end{figure}
Figure \ref{fig:ST1}a shows that slug length increases as the surface tension rises. Interestingly, PAM 0.1 and 0.2 wt\% solutions display almost similar behaviour. However, a noticeable difference is observed in case of PAM 0.4 wt\% solution. This is ascribed to the dominance of shear force exerted by the higher concentration solution over surface tension force during the rupture step. Slug pinch-off process accelerates at lower surface tension due to weak attraction forces between two-phases. Gas slug volume and unit cell length are also found to increase with surface tension as shown in Figure \ref{fig:ST1}b and c, respectively. Figure \ref{fig:ST1}d depicts that for all concentrations of PAM solution, pressure drop increases with increasing surface tension while a significant difference is observed between higher and lower concentration solutions, which is solely due to frictional pressure drop exerted by liquid having different effective viscosity values. 

\subsection*{Effect of contact angle}
\begin{figure}[h]
	\centering
	\includegraphics[width=0.8\textwidth]{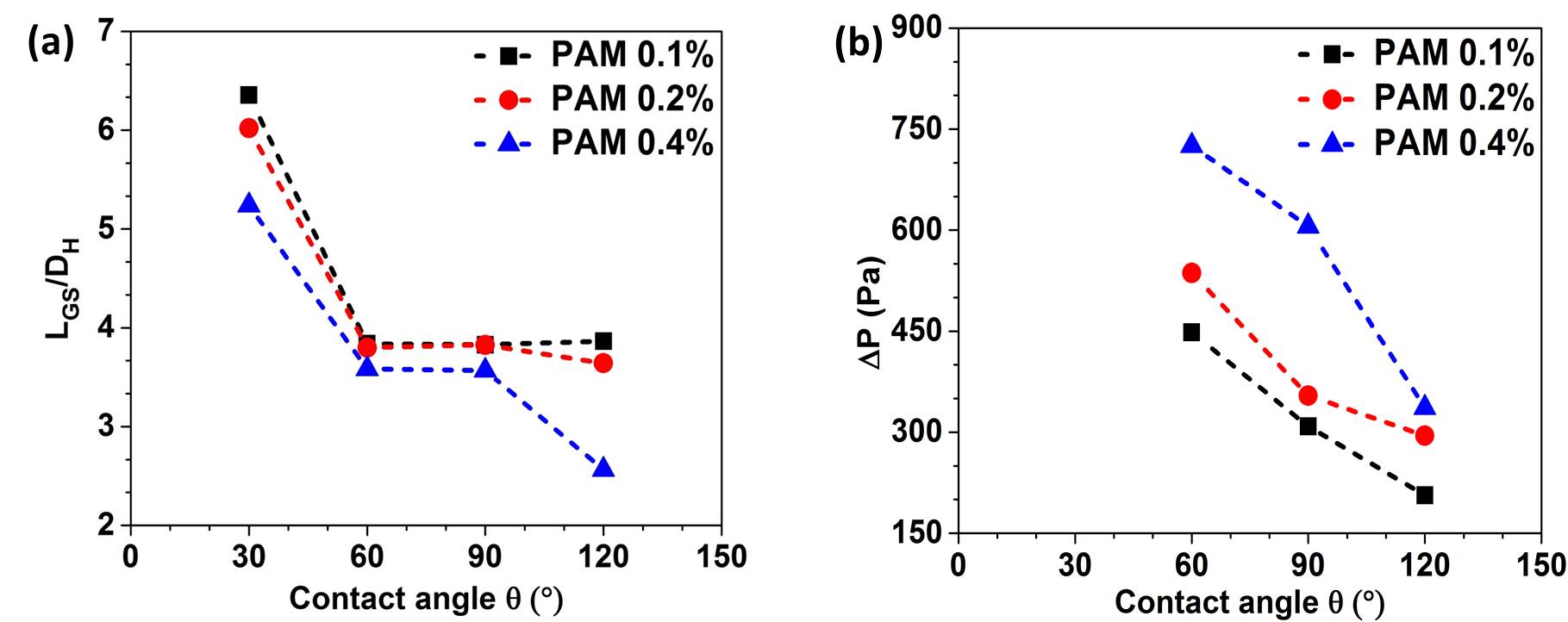}
	\caption{\label{fig:CA1} Effect of contact angle ($\theta$) on (a) gas slug length, (b) gas slug volume, (c) unit cell length, and (d) pressure drop at $J_G$= 0.1085 m/s, and $J_L$= 0.1085 m/s. }
\end{figure}

Influence of contact angle is investigated by altering it from 30 to 120\textdegree for all PAM solutions, and the results are presented in Figure \ref{fig:CA1}. It is observed that the gas slug length diminishes with increasing contact angle, although appreciable reduction is observed for solutions with higher PAM concentration (Figure \ref{fig:CA1}a). 
\begin{figure}[!ht]
	\centering
	\includegraphics[width=0.8\textwidth]{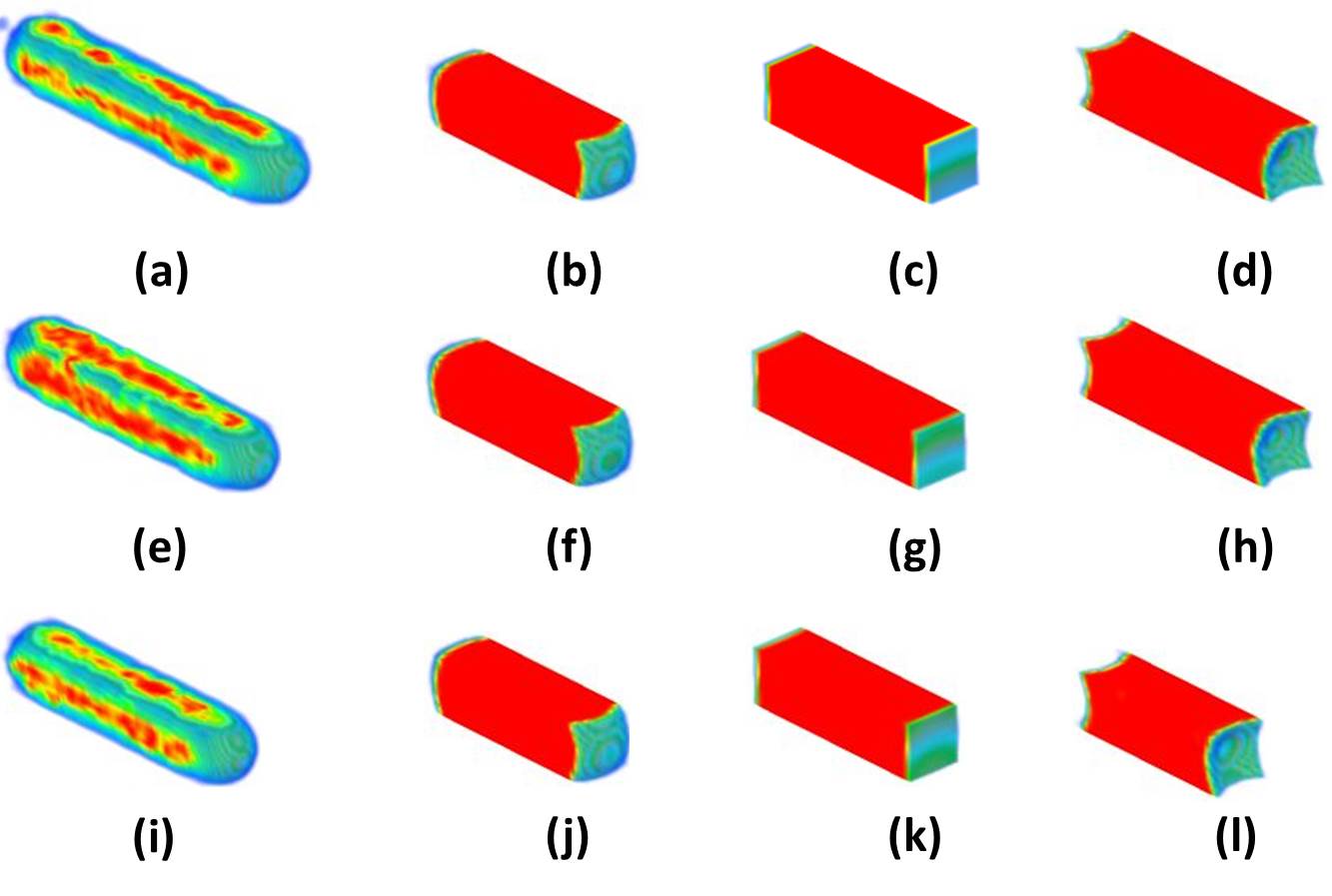}
	\caption{\label{fig:CA2} Effect of contact angle ($\theta$) on gas slug shape for PAM 0.1 wt\% (a) 30\textdegree, (b) 60\textdegree, (c) 90\textdegree, (d) 120\textdegree; for PAM 0.2 wt\% (e) 30\textdegree, (f) 60\textdegree, (g) 90\textdegree, (h) 120\textdegree; and for PAM 0.4 wt\% (i) 30\textdegree, (j) 60\textdegree, (k) 90\textdegree, (l) 120\textdegree at $J_G$= 0.1085 m/s, and $J_L$= 0.1085 m/s. }
\end{figure}
Pressure drop in gas\textendash liquid system strongly depends on frictional resistances of the liquid phase and rapid changes in the shape of gas slug.\cite{bretherton-1961,yue-2009} It is important to note that considerable difference is observed in gas slug shape from convex to concave with increasing contact angle. Figure \ref{fig:CA2} illustrates that both nose and rear of the gas slug change from convex to concave shape, which ultimately leads to a reduction in slug length. Consequently, the pressure drop is found to decrease for all PAM solutions with increasing contact angle, as shown in Figure \ref{fig:CA1}b. It is interesting to note that when the channel wall is considered to be hydrophobic and the gas phase is in direct contact with the microchannel wall then at larger contact angles, there tends to be no liquid film around the gas slugs, which is in agreement with the observations reported in the literature.\cite{cubaud-2006}

\section*{Conclusions}
The formation of segmented flow in a converging square microchannel is analysed for two-phase gas\textendash shear\textendash thinning liquid using CLSVOF method. The effect of PAM concentration, gas-liquid flow ratios, surface tension, and contact angle on gas slug length, gas slug volume, unit cell length, and pressure drop are demonstrated. It is observed that higher concentration of PAM solutions result in early detachment of gas slug, reduced gas slug length and volume. Furthermore, velocity profiles are also examined in different PAM concentrations. It is found that the velocity profiles are flatter at higher PAM concentrations, which indicate shear-thinning nature of the flow. Rheological properties of the solution are found to influence the formation of segmented flow, and their effects are more pronounced at higher concentration of PAM solutions. Gas slug length and volume are found to increase with $J_G$/$J_L$ ratio, and are considerably larger at higher gas velocities. Pressure drop increases with increasing $J_L$, which is substantial for solutions with higher PAM concentrations due to greater apparent viscosity. Due to the change in gas slug shape from convex to concave with increasing contact angle, reduction in pressure drop is observed. 


\section*{Nomenclature}
$Ca^{*}$= Modified Capillary number ($KU_B^n{D_H}^{(1-n)}/\sigma$) \\
\textit{a} = interface thickness (m)\\
$D_H$ = hydraulic diameter (m)\\
$U$  =  velocity (m/s) \\
$p$ = pressure (Pa)\\
$J_G$ = gas velocity (m/s)\\
$J_L$ = liquid velocity (m/s)\\
$J_{TP}$ = two phase velocity (m/s)\\
$L_{GS}$ = length of gas slug (m)\\
$L_{LS}$ = length of liquid slug (m)\\
$L_{UC}$ = length of unit cell (m)\\
$V_{GS}$ = volume of gas slug ($m^{3}$)\\
$n$ = power law index\\
$K$ = consistency index ($Pa.s^{n}$)\\
$t$ = flow time (s)\\
$\bigtriangleup$P =  pressure drop (Pa)\\
$\vec{U }$ = velocity vector (m/s)\\
$\vec{ F}_{SF} $ = volmetric surface tension force (N/$m^{3}$)\\
$\textit{H}(\varphi)$ =  Heaviside function\\
$W$ = width ($m$) \\
$\hat{N}$ = unit vector normal to wall\\
$\hat{M}$ = unit vector tangential to wall\\
\textit{Greek symbol}\\	
$\alpha$  = volume fraction\\
$\dot{\gamma } $ = shear rate (1/s)\\
$\delta$ = liquid film thickness (m) \\ 
$\theta$ = contact angle (\textdegree)\\
$\eta_{app}$= apparent viscosity ($Pa.s^{n}$)\\
$\rho$  = density (kg/$m^{3}$)\\
$\sigma$  = surface tension (N/m)\\ 
$\tau$ = shear stress (Pa)\\
$\kappa_{n}$ = radius of curvature (1/m)\\
$\kappa ( \varphi )$ = interface curvature\\
$\delta ( \varphi )$ = Direct delta function\\
$\varphi $ = level function\\
$\vec{\chi}$ = position vector\\
\textit{Subscripts}\\
$LS$ = liquid slug \\
$GS$ = gas slug\\
$G$ = gas\\
$L$ = liquid\\
$UC$= unit cell\\
$q$ =  $q^{th}$ phase\\

\section*{References}
\bibliography{Pankaj1}

\end{document}